\documentclass[a4paper,12pt]{article}
\usepackage{amsmath}
\usepackage{ulem}
\usepackage{calc}
\usepackage{vmargin}
\usepackage{amstext}
\usepackage{graphicx}
\usepackage{amsfonts}
\usepackage{amssymb}
\usepackage{indentfirst}

\begin{document}

	\title{
		{\bf Symmetries of linear and nonlinear partial differential equations. } }
	
	\author{ \bf O. V. Kaptsov
		\\ Institute of Computational Modelling SB RAS,
		\\ Krasnoyarsk, Russian Federation
		\\ E-mail: profkap@gmail.com}
	
	\date{}
	\maketitle
	
	\noindent
{\bf Abstract.}
We consider  higher symmetries and operator symmetries of linear partial differential equations. The higher symmetries form a Lie algebra, and operator ones form an associative algebra.
The relationship between these symmetries is established.
We show that symmetries of  linear equations  sometimes generate symmetries of  nonlinear ones. New symmetries of two-dimensional stationary equations of gas dynamics are found.


\section{Introduction}

It is well known that symmetries play a crucial role in finding solutions of differential equations. The theory of point symmetries is well described in numerous monographs and textbooks \cite{Ovs,Ibragimov,Olver}.
A large number of examples of invariant and partially invariant solutions are presented in various handbooks \cite{Ibr,Pol}. 
One can say that the theory of point transformations is very well developed.
 Some generalizations of the Lie theory have been proposed at present.
The most successful advances include the theory of higher symmetries of nonlinear equations and operator symmetries of linear equations \cite{Ibragimov,Vin,Miller,Kalnins}.
  However, the use of higher symmetry operators is complicated by the fact that the transformations constructed using these operators act in infinite-dimensional spaces and are represented by formal series \cite{Ibragimov}. As a result, it is difficult to determine analogs of invariant solutions with respect to such transformations.
  
  In this paper we introduce modified definitions of admitted operators and operator symmetries for linear systems of partial differential equations. It is easily verified that operator symmetries form an associative algebra with respect to ordinary multiplication and a Lie algebra with commutator multiplication. It is proved that to each operator symmetry there corresponds some admitted one. 
  It turns out that symmetries of linear equations can be transformed into symmetries of nonlinear equations in some cases. 
  Here, as an example, we consider a system of two equations,
  describing plane, steady, irrotational gas flows \cite{OlvGD,DAU}.
  After the  hodograph transformation a linear system is obtained. The admitted operators of this system give rise to an infinite series of symmetries of nonlinear gas dynamics equations.

\section{Symmetries}

Consider the matrix differential operator
\begin{equation} \label{LOp}
	L = \sum_{|\alpha|\geq 0}^{k} A_\alpha(x) \frac{\partial^{|\alpha|} }{\partial x_1^{\alpha_1}\cdots\partial x_n ^{\alpha_n}} ,
\end{equation}
where $\alpha= (\alpha_1,\dots,\alpha_n) $, $A_\alpha$ are $m\times m$ matrices depending on $x=(x_1,\dots,x_n)$.
The operator defines a system of linear partial differential equations
\begin{equation} \label{Lu=0}
	Lu = 0, \qquad 
\end{equation}
where $u=(u^1,\dots,u^m)$ is a set of unknown functions of $x$.

Further, the operator of total derivative \cite{Ovs,Ibragimov} with respect to $x_i$ 
is denoted by $D_{x_i}$. The expression $D^\alpha$ means the product of operators 
$D^{\alpha_1}_{x_1}\cdots D^{\alpha_n}_{x_n}$.

\noindent
{\bf Definition 1.} The system (\ref{Lu=0}) admits the operator
in canonical form
\begin{equation} \label{X}
	X = \sum_{j=1}^{m}\eta_j \frac{\partial }{\partial u^j} +
	\sum_{\substack{1\leq j\leq m \\ \alpha\in \mathbb{N}^n }} D^\alpha\eta_j 
	\frac{\partial }{\partial u^j_\alpha} , 	
\end{equation}
if there is a matrix differential operator $M$ such that the following equality is satisfied  
\begin{equation} \label{Sym}
	L\eta = MLu, 
\end{equation}
where  $u=(u^1,\dots,u^m)$ is a set of arbitrary smooth functions of $x$, and $\eta = (\eta_1,\dots,\eta_m)$. The relation (\ref{Sym}) will be called the defining equation.

The above definition differs from the standard one \cite{Ibragimov,Olver}. Obviously, the condition (\ref{Sym}) is sufficient for the classical invariance of the system (\ref{Lu=0}). It can be shown that it is necessary, but we will not need it. 

\noindent
{\it Remark.} If the system of partial differential equations $L(u)=0$ is nonlinear, then the condition (\ref{Sym}) must be replaced by the following
$$ XL(u) = ML(u) . $$

It follows from the formula (\ref{Sym}) that if $u$ is a solution of the system (\ref{Lu=0}), then $\tilde{u} =\eta$ is also a solution of this system. Thus the transformation
$$ x\longrightarrow x ,\qquad u\longrightarrow \eta $$
acts on the solutions of the system (\ref{Lu=0}). We will call such transformations $L$-symmetries.

{\bf Proposition 1.} If $\eta^1,...,\eta^p$ are solutions of the defining  equations 
\begin{equation} \label{Mk}
	L\eta^k = M_kLu, \qquad k=1,\dots p ,
\end{equation}
then 
\begin{equation} \label{Sum}
	x\longrightarrow x ,\quad u\longrightarrow \sum_{k=1}^p c_k\eta^k, \qquad c_k\in\mathbb{R} 
\end{equation}
is the $L$-symmetry of the equation (\ref{Lu=0}).

\noindent
Indeed, since the functions $\eta^k$ satisfy (\ref{Mk}), then, due to the linearity of the operators, the equality is true
$$ L(\sum_{k=1}^p c_k\eta^k) = (\sum_{k=1}^p c_kM_k)Lu .$$
This means that the transformation (\ref{Sum}) is $L$-symmetry.

\noindent
{\bf  Proposition 2.} The set of $L$-symmetries of the system (\ref{Lu=0}) forms a monoid with the composition operation.

This immediately follows from the fact that symmetries act on solutions of the system and therefore the composition of two $L$-symmetries of the system (\ref{Lu=0}) is a $L$-symmetry. Moreover, the identity transformation is also a symmetry.

The second method of introducing symmetries of linear equations is described in \cite{Miller, Kalnins}.
We provide modified version of definition.

\noindent
{\bf  Definition 2.} Let a differential operator (\ref{LOp}) be given. 
The differential operator $S$ is called the operator symmetry of the equation (\ref{Lu=0}),
if there is a differential operator $\mathcal{D}$ such that
\begin{equation} \label{Sym2}
	LS = \mathcal{D}L .
\end{equation}
It is assumed that $S$ is not a polynomial in $L$.

Obviously the operator symmetry $S$ acts on solutions of the equation (\ref{Lu=0}), i.e., transforms solutions into ones. 

\noindent
{\bf  Proposition 3}. Let $\mathcal{S}_1, \mathcal{S}_2$ be two operator symmetries of the system (\ref{Lu=0}). Then 
$$ b_1\mathcal{S}_1 + b_2\mathcal{S}_2, \quad \mathcal{S}_1 \mathcal{S}_2, \quad 
\mathcal{S}_1 \mathcal{S}_2 - \mathcal{S}_2 \mathcal{S}_1, $$
also operator symmetries for any $b_1, b_2\in\mathbb{R}$.

\noindent
{\it Proof}. By condition the following equalities are satisfied
$$ L \mathcal{S}_1 = \mathcal{D}_1 L , \qquad L \mathcal{S}_2 = \mathcal{D}_2 L .           $$
It follows that
$$ L (b_1\mathcal{S}_1 + b_2\mathcal{S}_2)  = b_1 L \mathcal{S}_1 +  b_2 L \mathcal{S}_2 =
b_1 \mathcal{D}_1 L +  b_2 \mathcal{D}_2 L = (b_1 \mathcal{D}_1  +  b_2 \mathcal{D}_2) L , $$
$$ L\mathcal{S}_1\mathcal{S}_2 =  \mathcal{D}_1 L\mathcal{S}_2 =\mathcal{D}_1\mathcal{D}_2 L ,$$
$$  L(\mathcal{S}_1\mathcal{S}_2 -\mathcal{S}_2\mathcal{S}_1) = \mathcal{D}_1\mathcal{D}_2 L -
\mathcal{D}_2\mathcal{D}_1 L = (\mathcal{D}_1\mathcal{D}_2  -
\mathcal{D}_2\mathcal{D}_1)L .$$

\noindent
{\it Remark.} If we introduce the commutator of operators $\mathcal{S}_1, \mathcal{S}_2$ according to the well-known formula
$[\mathcal{S}_1, \mathcal{S}_2]= \mathcal{S}_1\mathcal{S}_2 -\mathcal{S}_2\mathcal{S}_1$, then the last expression in the proof is rewritten as
$$ L[\mathcal{S}_1, \mathcal{S}_2]=[\mathcal{D}_1, \mathcal{D}_2]L .$$

\noindent
{\bf Corollary.} The set of operator symmetries of the system (\ref{Lu=0}) forms an associative algebra over $\mathbb{R}$ with usual multiplication and a Lie algebra with commutator one.

\noindent
{\bf Proposition 4}. If $S$ is an operator symmetry of the equation (\ref{Lu=0}),
$u=(u^1,\dots,u^m)$ is a set of smooth functions, then 
$\eta = Su $ is a solution to the defining equation that generates $L$-symmetry.

By assumption, there exists an operator $\mathcal{D}$ satisfying the equality
(\ref{Sym2}). Applying to $u$ the operators on the left and right sides (\ref{Sym2}),
we obtain the equality (\ref{Sym}), in which $\eta = Su $, and $M=\mathcal{D}$.

\section{Symmetries of  stationary  gas dynamics equations }

The $L$-symmetries introduced above can be applied to some nonlinear equations.
As an example, consider the well-known system of stationary equations  
\begin{equation} \label{GD} 
	u_y - v_x=0, \quad (u^2-c^2)u_x + 2uvu_y +(v^2-c^2)v_y =0 ,
\end{equation}
describing isentropic irrotational steady  plane flows of compressible fluid \cite{OlvGD,DAU}.
Here $u, v$ are the components of the velocity vector, $c$ is the speed of sound, expressed from the Bernoulli integral
$$ u^2 +v^2 + I(c^2) = const .$$

After passing to the hodograph variables, we obtain a system of linear equations \cite{OlvGD}
\begin{equation} \label{God} 
	x_v - y_u=0, \quad (u^2-c^2)y_v - 2uvx_v +(v^2-c^2)x_u =0 ,
\end{equation}
for two unknown functions $x, y$ depending on $u,v$.
It is not difficult to see \cite{OlvGD} that both systems admit the  rotation, scaling and translation operators
$$ -v\frac{\partial }{\partial u} + u\frac{\partial }{\partial v} - y\frac{\partial }{\partial x}
+ x\frac{\partial }{\partial y}\ , \quad x\frac{\partial }{\partial x} + y\frac{\partial }{\partial y}\ ,\quad  \frac{\partial }{\partial x}\ , \quad  \frac{\partial }{\partial y}\ .$$

The rotation operator admitted by the system (\ref{God}) in canonical form is expressed by
$$ (-y+vx_u -ux_v)\frac{\partial }{\partial x} + (x+vy_u-uy_v) \frac{\partial }{\partial y}\ . $$
Therefore, according to Proposition 1, the transformation
$$ \tilde{u} = u\ ,\quad \tilde{v} = v\ ,\quad \tilde{x} = -y+v x_u-u x_v \ ,
\quad \tilde{y} = x+v y_u-u y_v $$
acts on solutions of a linear system (\ref{God}) and is an $L$-symmetry of this system.
Using three other symmetry operators, we obtain $L$-symmetry of the form 
\begin{gather} 
	\tilde{u} = u ,\quad \tilde{v} = v \label{E1}\\
	\tilde{x} = c_1(-y+v x_u-u x_v) +c_2x +c_3 ,
	\quad \tilde{y} = c_1(x+v y_u-u y_v) + c_2y +c_4 ,\label{E2}
\end{gather}
where $c_i$ are arbitrary constants.

In order to obtain the symmetries of the gas dynamics equations (\ref{GD}), it is necessary to express the derivatives $x_u, x_v, y_u, y_v$ in terms of the derivatives $u_x, u_y, v_x, v_y$.
 Using the hodograph transformation, it is easy to find these derivatives \cite{OlvGD}
 $$ x_u = v_y/J ,\quad x_v= - u_y/J, ,\quad y_u = -v_x/J ,\quad y_v= u_x/J , $$
 where $J=\frac{\partial (u,v)}{\partial (x,y)}$ is the Jacobian of the functions $u , v$.
 Thus, the formulas of transformation (\ref{E1}), (\ref{E2}) are rewritten  in the form
\begin{gather} 
	\tilde{u} = u ,\quad  \tilde{v} = v   \label{E3} \\ 
	\tilde{x} = c_1\left(-y+\frac{vv_y+u u_y}{J}\right) +c_2x +c_3   ,
	\quad  \tilde{y} = c_1\left(x -\frac{vv_x+uu_x}{J}\right) + c_2y +c_4 \label{E4}.
\end{gather}
The last formulas determine the transformation of solutions of the system (\ref{GD}) back into solutions of this system.

Composition of transform (\ref{E1}), (\ref{E2}) and 
\begin{gather} 
	\hat{u} = \tilde{u} ,\quad \hat{v} = \tilde{v} \nonumber\\
	\hat{x} = b_1(-\tilde{y} +\tilde{v} \tilde{x}_u -\tilde{u}\tilde{x}_v) 
	+b_2\tilde{x} +b_3 ,
	\quad \hat{y} = c_1(\tilde{x} +\tilde{v} \tilde{y}_u-\tilde{u} \tilde{y}_v) 
	+ b_2\tilde{y} +b_4 , \nonumber
\end{gather}
gives a new second-order symmetry of the system (\ref{God}).
   One can obtain symmetries of the system (\ref{God}) of arbitrary order by means of compositions.
  Thus, an infinite series of symmetries of the system in the hodograph variables arises. An infinite series of symmetries of the gas dynamics equations (\ref{GD}) are obtained by recalculating the corresponding derivatives.

\section{Conclusion}
Using the found symmetries, it is possible to construct solutions from known ones. For example, if we take scale-invariant  solutions \cite{OlvGD} of the system (\ref{GD}), then formulas (\ref{E3}), (\ref{E4}) will generate new solutions.
The first articles devoted to new types of symmetries have appeared recently \cite{Kap1,Kap2}. This approach requires further development and construction of more new examples. 

This work is supported by the Krasnoyarsk Mathematical Center and financed by the Ministry of Science and Higher Education of the Russian Federation in the framework of the establishment and development of regional Centers for Mathematics Research and Education (Agreement No. 075-02-2024-1378).

\end{document}